\documentclass[%
 reprint,
nofootinbib,
 aps,
 onecolumn,
]{revtex4}

\usepackage{amssymb,amsmath,amsfonts,amsbsy,graphicx}
\usepackage{color}

\newcommand\be{\begin{equation}}
\newcommand\ee{\end{equation}}
\newcommand\ba{\begin{eqnarray}}
\newcommand\ea{\end{eqnarray}}\newcommand\eq{\begin{equation}}
\newcommand\en{\end{equation}}

\usepackage{bm}
\usepackage{subfig}
\usepackage{url}

\def\gsim{\;\rlap{\lower 2.5pt
 \hbox{$\sim$}}\raise 1.5pt\hbox{$>$}\;}
\def\lsim{\;\rlap{\lower 2.5pt
 \hbox{$\sim$}}\raise 1.5pt\hbox{$<$}\;}

\newcommand{\mnras}{Monthly Notices of the RAS}

\newcommand{\aap}{Astronomy \& Astrophysics}
\newcommand{\apjl}{ApJL}
\newcommand{\apjs}{ApJS}
\newcommand{\jcap}{JCAP}

\usepackage{ragged2e}
\DeclareCaptionJustification{justified}{\justifying}
\begin{document}
\title{
Constraining Mixed Dark-Matter Scenarios of WIMPs and Primordial Black Holes from CMB and 21-cm observations
}
\author{Hiroyuki Tashiro$^1$ and Kenji Kadota$^2$ \\
  {\small $^1$ Department of physics and astrophysics, Nagoya University, Nagoya 464-8602, Japan}\\
    {\small $^2$ Center for Theoretical Physics of the Universe, Institute for Basic Science (IBS), Daejeon, 34051, Korea}
  }


\begin{abstract}
We consider the dark matter (DM) scenarios consisting of the mixture of WIMPs and PBHs and study what fraction of the total DM can be PBHs.
In such scenarios, PBHs can accrete the WIMPs and consequently enhance the heating and ionization in the intergalactic medium due to WIMP annihilations.
We demonstrate that the CMB data can give the stringent bounds on the allowed PBH fraction which are comparable or even tighter than those from the gamma-ray data depending on the DM masses.
For instance, the MCMC likelihood analysis using the Planck CMB data leads to a bound on the PBH DM fraction with respect to the total dark matter $f_{\rm PBH} \lesssim {\cal O}( 10^{-10}\sim 10^{-8})$ for the WIMP mass $m_{\chi}\sim {\cal O}(10\sim 10^3)$ GeV with the conventional DM annihilation cross section $\langle \sigma v \rangle=3 \times 10^{-26}~\rm cm^3/s $.
We also investigate the feasibility of
the global 21-cm signal measurement to provide the stringent constraints on the PBH fraction.
\end{abstract}

\maketitle

\setcounter{footnote}{0}
\setcounter{page}{1}\setcounter{section}{0} \setcounter{subsection}{0}
\setcounter{subsubsection}{0}

\section{Introduction}
There have been revived interests in the primordial black hole~(PBH) dark matter scenarios since the LIGO/Virgo detection of black-hole mergers \cite{TheLIGOScientific:2016pea,LIGOScientific:2018jsj}. While the parameter space for PBHs to account for the total dark matter component has been narrowed by many other complementary observation data such as those from gravitational lensing, the PBH can well be a partial dark matter component~\cite{Carr:2020xqk,Green:2020jor,Carr:2020gox}. For the dark matter~(DM) candidate accounting for the rest of the dark matter in the presence of such PBH partial dark matter, the widely discussed weakly interacting massive particles~(WIMPs) can be an intriguing possibility. Such WIMP-PBH mixed dark matter scenarios lead to ultracompact minihalos (UCMHs) around the PBHs (so-called "dressed PBHs"), and one can expect enhanced DM annihilation from those steep DM profiles around the PBHs. In the previous literature, many used the gamma-ray data such as those from Fermi Large Area Telescope (LAT) and pointed out the incompatibility of the coexistence of WIMPs and PBHs due to the current nonobservation of enhanced gamma-ray emission from those UCMHs \cite{Ricotti:2007jk,Mack:2006gz,Ricotti:2009bs,Gondolo:1999ef,Lacki:2010zf,Boucenna:2017ghj,Adamek:2019gns,Eroshenko:2016yve,Carr:2020mqm,Cai:2020fnq,Delos:2018ueo,Kohri:2014lza,Bertone:2019vsk,Ando:2015qda,Hertzberg:2020kpm,Yang:2020zcu,Zhang:2010cj}.
We note that the PBHs were produced in the radiation-dominated epoch and the dark matter could be gravitationally bound to PBHs to form the UCMHs around them by the CMB epoch $z\sim 10^3$ \cite{Ricotti:2007au,Ricotti:2007jk,Mack:2006gz,Adamek:2019gns,Carr:2020xqk,Green:2020jor,Carr:2020gox}.
One can then expect the inevitable effects of the energy injection from the DM annihilation on the CMB observables.
The radiation emitted by the gas falling onto the PBHs and their effects on the CMB (temperature/polarization anisotropy and spectral distortions) have been actively discussed so far~\cite{Ali-Haimoud:2016mbv, Bernal:2017vvn,2017PhRvD..96h3524P,2019PhRvD..99j3519A}, but the DM annihilation effects on the CMB in the presence of the dressed PBHs have been less explored \cite{Carr:2020xqk,Green:2020jor,Carr:2020gox}. For instance a pioneering work \cite{Ricotti:2007au} studied how the X-rays emitted by gas accretion onto PBHs can modify the CMB observables by including the effects of the DM accumulation around the PBHs on the gas accretion rate, but the DM annihilation was not considered in their Markov-Chain Monte Carlo (MCMC) likelihood analysis (see e.g. \cite{Ali-Haimoud:2016mbv,Bernal:2017vvn,2019PhRvD..99j3519A,Serpico:2020ehh,Carr:2020xqk,Green:2020jor,Carr:2020gox,Kohri:2014lza} for more recent relevant works). We in this paper assume the Majorana DM particles which can self-annihilate (a typical example is WIMP), and perform the MCMC likelihood analysis on the PBH abundance by studying how the energy injection by the DM annihilation from the dressed PBHs can affect the CMB observables due to the change in the thermal and ionization history of the Universe. Our CMB bounds on the allowed PBH fraction can give comparable or better bounds than those from gamma-ray observations depending on the DM mass. For instance, our MCMC analysis shows that the CMB bounds on PBH fraction $f_{\rm PBH}$ (with respect to the total DM) are $f_{\rm PBH} \lesssim 3\times 10^{-10},~7\times 10^{-9}$, and $3\times 10^{-8}$ for $m_{\chi}=$ 10~GeV, 100~GeV and 1~TeV, respectively, assuming the s-wave annihilation into $b\bar{b}$ with $\langle \sigma v \rangle =3 \times 10^{-26} \rm cm^3/s$ (to be compared with the corresponding Fermi gamma-ray data bounds $f_{\rm PBH} \lesssim 10^{-9},2\times 10^{-9}$, and $4\times 10^{-9}$ \cite{Adamek:2019gns,Ando:2015qda}).
In addition to the CMB bounds using the Planck data, we also present a brief discussion on the 21cm constraints using the EDGES result, which is also affected by the energy injection from DM annihilation \cite{2018Natur.555...67B}.

The paper is organized as follows. Section \ref{sec2} reviews the energy injection rate due to the DM annihilation from the steep DM profile around a PBH. Section \ref{sec3} discusses how such energy injection from DM annihilation can affect the thermal evolution of the intergalactic medium and performs the MCMC likelihood analysis using the Planck data leading to the stringent bounds on the allowed PBH fraction in the presence of WIMPs. We also discuss how the global 21-cm signals can be affected in the mixed PBH-WIMP dark matter scenarios. Section \ref{secconc} is devoted to the conclusion.

\section{WIMP annihilation from DM halo around PBHs}
\label{sec2}

\subsection{Dark matter halo profile}

In the mixed-DM scenarios consisting of PBHs and WIMPs,
PBHs can gravitatinally bound nonrelativistic WIMP DM particles soon after their formation and, as a result, dress a halo of WIMPs whose density profile is a spike type.
Here we construct the model of the DM halo around a PBH following Ref.~\cite{Adamek:2019gns}.

During the radiation-dominated epoch, the turnaround scale at a redshift~$z$, where the gravitational attraction from a PBH decouples a DM mass shell from the Hubble flow, is obtained numerically~\cite{Adamek:2019gns},
\be
r_{\rm ta} (z) \approx (2GM_{\rm PBH} t^2)^{1/3} ,
\ee
where $t$ is a time corresponding to a redshift~$z$.

We simply assume that, when the dark matter particles decouple from the Hubble flow, the DM density of each mass shell matches the background density at its turnaround time.
This simple assumption provides the steep density profile around a PBH, and the density profile
at the redshift of matter-radiation equality, $z_{\rm eq}$, is
\be
\label{eq:dmprofile}
\rho(r) = \overline \rho_{\rm DM}(z_{\rm eq}) \left( \frac{r}{r_{\rm ta} (z_{\rm eq})}\right)^{-9/4}
\quad {\rm for}~r<r_{\rm ta}(z_{\rm eq}) ,
\ee
where we assume that the DM fraction of the WIMP is almost unity, $f_{\chi} \approx 1$,
and
$\overline \rho_{\rm DM}(z) $ is the background DM density at a redshift $z$.
After the time of matter-radiation equality, the dark matter halo can grow
by the secondary infalling into the halo in Eq.~\eqref{eq:dmprofile}~\cite{1985ApJS...58...39B}.
Even in this stage, it is confirmed that
the profile shape in Eq.~\eqref{eq:dmprofile} is not disrupted and the infalling matter creates the NFW profile at large radii~\cite{Adamek:2019gns}.
Therefore, we assume that the density profile in Eq.~\eqref{eq:dmprofile} is valid
even at $z < z_{\rm eq}$.

The WIMP can annihilate away, in particular, in the central region of a halo because of its high density and soften the density cusp.
The maximum possible WIMP density in the halos is evaluated as~\cite{1992PhLB..294..221B} 
\be
\rho_{\rm max} (z) \approx \frac{m_\chi}{\langle \sigma v \rangle t} .
\ee

This maximum density gives the flat DM density inner region of a halo extending to $r_{\rm cut}$, where the power-law profile starts.
Considering the profile in Eq.~\eqref{eq:dmprofile}, we obtain
\be
r_{\rm cut} (z) = \left( \frac{\rho _{\rm max} (z)}{ \overline \rho_{\rm DM}(z_{\rm eq})} \right)^{-4/9} r_{\rm ta}(z_{\rm eq}).
\ee
We can hence model the dark matter density profile around a PBH at a redshift $z$ as
\be
\label{eq:dmprofile_z}
\rho(z, r) =
 \begin{cases}
    \rho_{\rm max}(z) \quad &{\rm for}~r < r_{\rm cut}(z)  \\
    \rho_{\rm max}(z)  (r/ r_{\rm cut}(z))^{-9/4}\quad &{\rm for}~r_{\rm cut}(z) \leq r < r_{\rm ta}(z_{\rm eq}) . \\
  \end{cases}
\ee

We focus on the scenarios where the DM kinetic energy is negligible compared with the potential energy around a PBH, for which the above profile $\rho(r)\propto r^{-9/4}$ is numerically verified~\cite{Adamek:2019gns,Serpico:2020ehh}.
One can neglect the DM kinetic energy when the potential energy dominates,
at least, at $r_{\rm cut} (z_{\rm eq})$ satisfying~\cite{Adamek:2019gns}
\be
\frac{E_{\rm k}}{E_{\rm p}}
=\left(\frac{T_{\rm KD}}{m_{ \chi}}\right) \left(\frac{t_{\rm KD}}{t_{\rm eq }}\right)
\left(\frac{\rho_{\rm max}(z_{\rm eq})}{ \overline \rho_{ \rm DM} (z_{\rm eq})}\right)^{2/9}
\left(\frac{r_{\rm ta}(z_{\rm eq})}{GM_{\rm PBH}} \right) \lesssim 0.01.
\ee
This equation
leads to the relation,
\be
M_{\rm PBH} \gtrsim 6.5 \times 10^{-4}~M_\odot ~ \left(\frac{\langle \sigma v \rangle }{3 \times 10^{-26} \rm cm^3/s }\right)^{-1/3} \left(\frac{m_\chi}{ 10~{\rm GeV }}\right)^{-73/24}.
\label{eq:Mpbh_mchi}
\ee
The cases when the effect of DM kinetic energy is non-negligible can be heavily dependent on the nature of DM kinetic decoupling and are left for future work \cite{Loeb:2005pm,Bertschinger:2006nq,Gondolo:2012vh, Profumo:2006bv,Gondolo:2016mrz,Green:2003un,Green:2005fa,Bringmann:2006mu,Eroshenko:2016yve,Boucenna:2017ghj}.

\subsection{Energy injection of WIMP annihilation}

The rate of DM annihilation is proportional to the density squared. Using the dark matter halo profile in Eq.~\eqref{eq:dmprofile_z}, the injected energy from a halo around a PBH into the cosmic plasma is given by
\ba
\left. \frac{d \cal E}{dt}
\right|_{\rm single}&&=
f_{\rm ann} (z) m_{\chi} \langle \sigma v \rangle
\int _{V_{\rm halo}} dV \left( \frac{\rho (r)}{m_\chi} \right)^2
\\ \nonumber
&&\approx
{4 \pi } f_{\rm ann} (z)\frac{ \langle \sigma v \rangle}{m_{\chi}}
\rho_{\rm max}^2(z) r_{\rm cut}^3 (z)
= {3} M_{\rm PBH}f_{\rm ann} (z)  \frac{ \Omega_{\rm DM}}{\Omega_{\rm M}}
\left[  \frac{ \langle \sigma v \rangle}{m_{\chi}} H^2(z) \overline \rho_{\rm DM}(z_{\rm eq}) \right]^{1/3}
,
\ea
where $V_{\rm halo}$ is the volume of a dark matter halo whose size is given by the turnaround radius at each redshift.
In the equation, $f_{\rm ann}(z)$ represents the fraction of the annihilation energy absorbed
by the plasma in the on-the-spot approximation.
The function $f_{\rm ann}(z)$ depends on the WIMP mass and the annihilation channel, and although $f_{\rm ann}(z)$ also depends on the redshift, we use the constant $f_{\rm ann}(z) = f_{\rm ann}$ hereafter ~\cite{Galli:2013dna}).
Now, we assume that PBHs have
a monochromatic mass function with the mass $M_{\rm PBH}$,
and the energy density fraction of PBHs
to the total dark matter is $f_{\rm PBH}$ (the current data require $f_{\rm PBH } \ll 1$, as discussed in the next section).
The number density of PBHs at a redshift~$z$ is
\be
n_{\rm PBH}(z) =  \frac{f_{\rm PBH}}{M_{\rm PBH}}\overline \rho_{\rm DM} (z).
\ee
Accordingly, the injected energy density due to the DM annihilation from total halos is
\be
\label{eq:injection_pbh}
\left. \frac{d^2E}{dVdt} \right\vert_{\rm PBHs} =
n_{\rm PBH}(z) \left. \frac{d \cal E}{dt}
\right|_{\rm single}
\propto
\left( \frac{ \langle \sigma v \rangle}{m_{\chi}} \right)^{1/3} (1+z)^4 f_{\rm ann}f_{\rm PBH}
,
\ee
where, in order to obtain the last expression,
we assume $t \approx 1/H(z)$ and the matter-dominated epoch.
Note that the injected energy density does not depend on $M_{\rm PHB}$ because $n_{\rm PBH}(z) \propto M_{\rm PBH}^{-1}$ and ${d \cal E}/{dt} \propto M_{\rm PBH}$.

The smooth-background DM also contributes to the energy injection of DM annihilation.
Therefore,
the total energy injection rate per volume due to the DM annihilation is
\be
\label{eq:dEdVdt}
 \frac{d^2E}{dVdt} (z) = \left. \frac{d^2E}{dVdt} \right\vert_{\rm sm}+\left. \frac{d^2E}{dVdt} \right\vert_{\rm PBHs}.
\ee
Here, the subscript "sm" denotes the smooth-background contributions,
\be
\label{eq:dEdVdt_smooth}
\left. \frac{d^2E}{dVdt} \right\vert_{\rm sm}= f_{\rm ann}\frac{\langle \sigma v \rangle}{ m_{\chi}}
 \overline \rho_{\rm DM}^2(z).
\ee

It would be useful to introduce the boost factor, $B(z)$, which represents the ratio of the PBH contribution to the smooth-background one,
\be
\label{eq:boost}
B(z) = \left. \frac{d^2E}{dVdt} \right\vert_{\rm PBHs} {\bigg/}  \left. \frac{d^2E}{dVdt} \right\vert_{\rm sm}
\approx 0.74 \times
\left( \frac{ \langle \sigma v \rangle/m_{\chi}}{
3 \times 10^{-27} {\rm cm^3/s/GeV}} \right)^{-2/3} \left( \frac{1+z}{100} \right)^{-2} \left( \frac{f_{\rm PBH}}{10^{-10}}\right).
\ee
The boost factor tells us that the contribution of PBHs relatively increases at lower redshifts.
Currently the measurement of CMB anisotropy provides the bound, $f_{\rm ann} \langle \sigma v \rangle /m_{\chi}\lesssim 3 \times 10^{-27} \rm cm^3 s^{-1}GeV^{-1}$ for the s-wave annihilating DM, considering only the smooth-background contribution~\cite{Slatyer:2015jla,Jungman:1995df}. According to Eq.~\eqref{eq:boost},
when $\langle \sigma v \rangle/m_{\chi} =
3 \times 10^{-27} {\rm cm^3/s/GeV}$, DM halos around PBHs with $f_{\rm PBH} \approx 10^{-10}$ can contribute
the modification of CMB anisotropy at the same level as the smooth-background DM at lower redshift~$z<100$.
One can hence infer that CMB observations could possibly provide a constraint on the PBH abundance $f_{\rm PBH} < {\cal O}(10^{-10})$ which could be competitive to the bounds from the gamma-ray observations.

\section{Constraints on the PBH abundance from cosmological observations}
\label{sec3}

The injected energy by the WIMP annihilation is absorbed into cosmic plasma through various channels including collisional heating and ionization.
Therefore, the injected energy can modify the evolution of the ionization fraction~$x_e$ and the baryon temperature.
Including the effect of the WIMP annihilation, we can write down the
evolution of the ionization fraction~\cite{Chen:2003gz},
\begin{equation}
(1+z)\frac{d x_e}{d z} = \frac{1}{H(z)}
\left[ R_s(z) - I_s(z) - I_x(z)  \right],
\label{eq:dxe}
\end{equation}
where $R_s$ and $I_s$ are the standard primordial hydrogen recombination
rate and ionization rate, respectively, and
$I_x$ represents the ionization contribution of the WIMP annihilation~\cite{1968ApJ...153....1P,1968ZhETF..55..278Z}.

The baryon temperature evolution in the presence of the WIMP annihilation can be obtained from
\begin{equation}
(1+z)\frac{dT_k}{dz}=2 T_k + \frac{8\sigma_T a_R T_{CMB}^4}{3m_e
c H(z)}\frac{x_e(T_k -T_{\gamma})}{(1+f_{\rm He}+x_e)}
 -\frac{2}{3
k_B H(z)} \frac{K_h(z)}{(1+f_{\rm He}+x_e)} ,
\label{eq:dT}
\end{equation}
where $K_h$ is the extra heating term by the DM annihilation.

The WIMP contributions, $I_x$ and $K_h$, are related to the energy injection given in
Eq~\eqref{eq:dEdVdt}\footnote{Although we neglect to discuss it in this paper, dark matter halos due to the standard scale-invariant adiavatic spectrum also can enhance the DM annihilation in particular, during the epoch of reionization~(for reference, see~\cite{2009A&A...505..999H}).},
\begin{eqnarray}
I_x=\frac{\chi_x} {{n_{\rm H}(z)}}\frac{d^2E}{dVdt} (z),
\nonumber \\
K_h=\frac{\chi_h} {E_i {n_{\rm H}(z)}}\frac{d^2E}{dVdt} (z),
\end{eqnarray}
where $E_i$ is the ionization energy of hydrogen, while
$\chi_x$ and $\chi_h$ provide the fractions of energy consumed for the ionization and heating.
These fractions mainly depend on the ionization fraction.
For example, in the neutral gas case,
it is known that roughly one third of the energy is used for the ionization,
another third goes into the excitation of gas atoms and
the rest is consumed for the heating~\cite{1985ApJ...298..268S}.
Here, for simplicity, we assume that the injected energy is quickly damped into the plasma and used for ionizing and heating, i.e.,~on-the-spot approximation and
we adopt the functions of $\chi_x$ and $\chi_h$ provided by Ref.~\cite{2017JCAP...03..043P}, which are the fitting formulas of the results in Ref.~\cite{Galli:2013dna}.

\begin{figure}
 \begin{tabular}{cc}
 \begin{minipage}{0.45\hsize}
  \begin{center}
   \includegraphics[width=75mm]{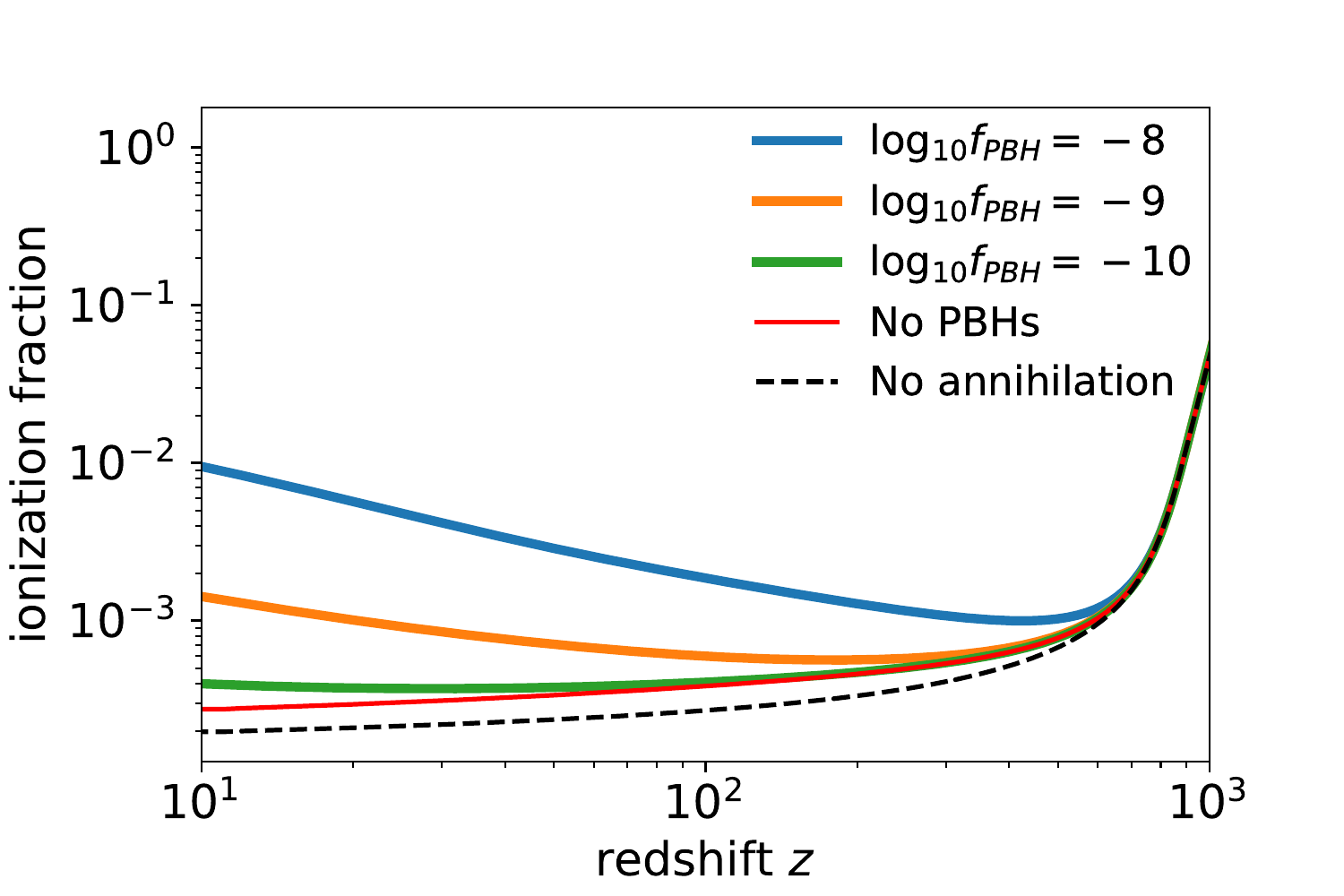}
  \end{center}
 \end{minipage}
 \begin{minipage}{0.45\hsize}
  \begin{center}
   \includegraphics[width=75mm]{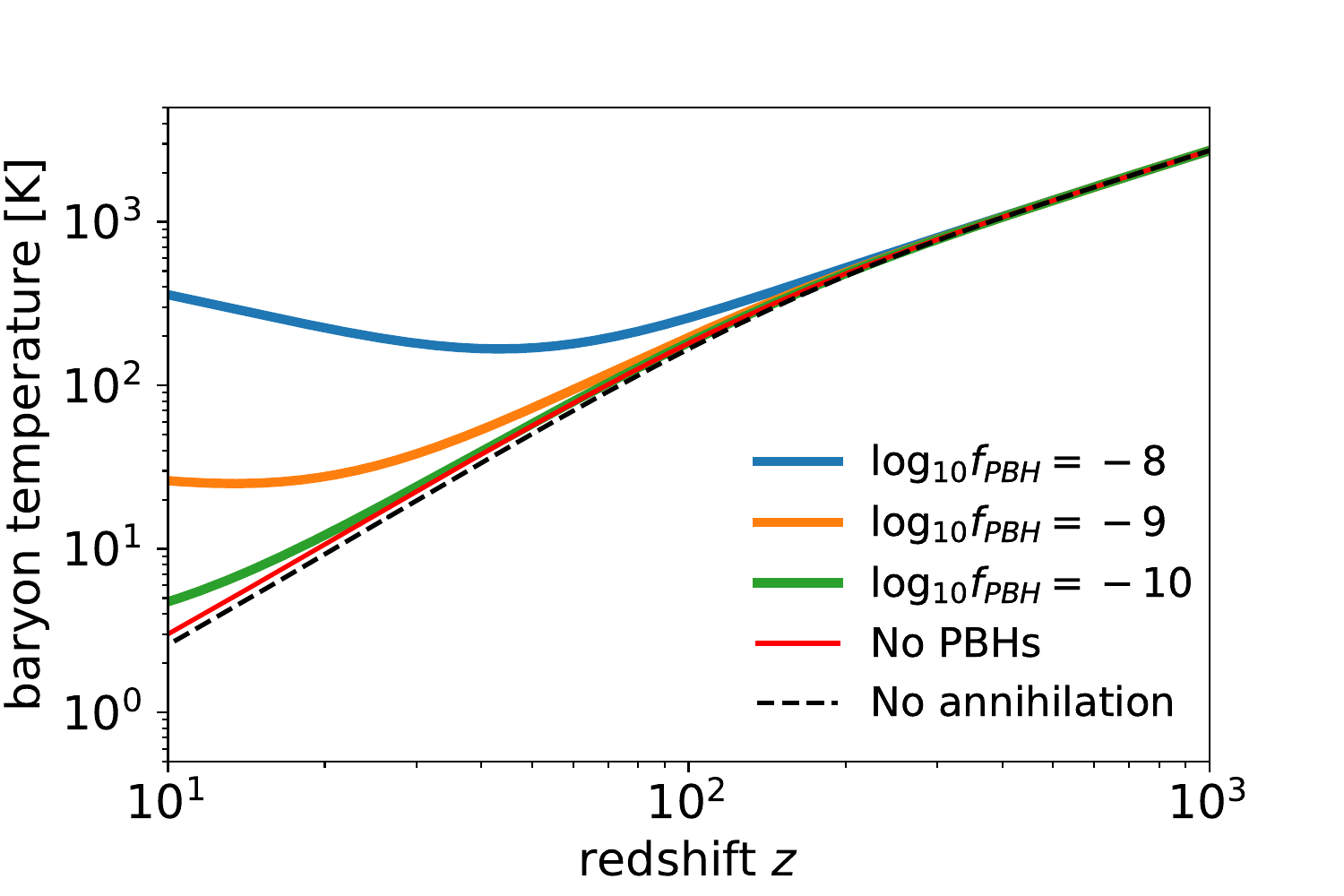}
  \end{center}
 \end{minipage}
 \end{tabular}
   \caption{The evolution of the ionization fraction~(left panel) and the baryon temperature~(right panel). We set $\langle \sigma v \rangle /m_{\chi} = 3 \times 10^{-28} \rm  cm^3 /s/ GeV$. From top to bottom, the solid lines represent the evolutions for $f_{\rm PBH} = 10^{-8}$, $f_{\rm PBH} = 10^{-9}$, $f_{\rm PBH} = 10^{-10}$ and $f_{\rm PBH} = 0$~(No PBHs). For reference, we plot the evolution without DM annihilation in dashed lines.
   }
  \label{fig:xe_temp}
\end{figure}

Including the primordial helium contribution, we obtain the ionization and
thermal histories of baryons from Eqs.~\eqref{eq:dxe} and \eqref{eq:dT} using {\tt HyRec}~\cite{2011PhRvD..83d3513A}.
Figure~\ref{fig:xe_temp} shows the ionization fraction (left panel) and the baryon temperature~(right panel) as functions of redshift.
Here we use $\langle \sigma v \rangle /m_{\chi} = 3 \times 10^{-28} \rm  cm^3 /s/ GeV$.
The current CMB bound (obtained without assuming PBHs) is $\langle \sigma v\rangle/m_{\chi}\lesssim 3 \times 10^{-27} \rm cm^3 s^{-1}GeV^{-1}$ for the s-wave annihilating DM and we conservatively use the values $\langle \sigma v\rangle/m_{\chi}\lesssim 3 \times 10^{-28} \rm cm^3 s^{-1}GeV^{-1}$ for the sake of illustration~\cite{Slatyer:2015jla}. Such a choice of $\langle \sigma v\rangle/m_{\chi}$ well below the current CMB bound ensures that there can be an enough room for the PBH contribution to dominate the smooth DM background contribution as inferred from Eq.~(\ref{eq:boost}).
In the figure, in order to show the dependence on $f_{\rm PBH}$,
the solid blue, orange, green, and red lines represent the evolutions for $f_{\rm PBH} = 10^{-8}$, $f_{\rm PBH} = 10^{-9}$, $f_{\rm PBH} = 10^{-10}$ and $f_{\rm PBH} = 0$, respectively.
For comparison, we plot the results
for the no-annihilation
case in a black dashed line.
As shown in Eq.~\eqref{eq:injection_pbh}
a large $f_{\rm PBH}$ increases the energy injection from dark matter halos around PBHs.
Therefore, a large $f_{\rm PBH}$ provides a strong impact on both the ionization fraction and the thermal evolution.
It
induces early reionization and heating of baryons.
Such modifications on these evolutions can be probed by cosmological observations including
CMB and redshifted 21-cm signals.
In the next section, we discuss the constraint on $f_{\rm PBH}$
from these observations.

\subsection{CMB anisotropy}

The existence of PBHs can enhance the WIMP annihilation and heat up and ionize baryons before the conventional epoch of reionization as shown in the previous section.
Since such early-epoch energy injection contributes to the increase in the optical depth of CMB from the last scattering surface,
the measurement of CMB anisotorpy can provide useful information on the PBH abundance with WIMP annihilation.

\begin{figure}
  \begin{center}
  \includegraphics[width=10.0cm]{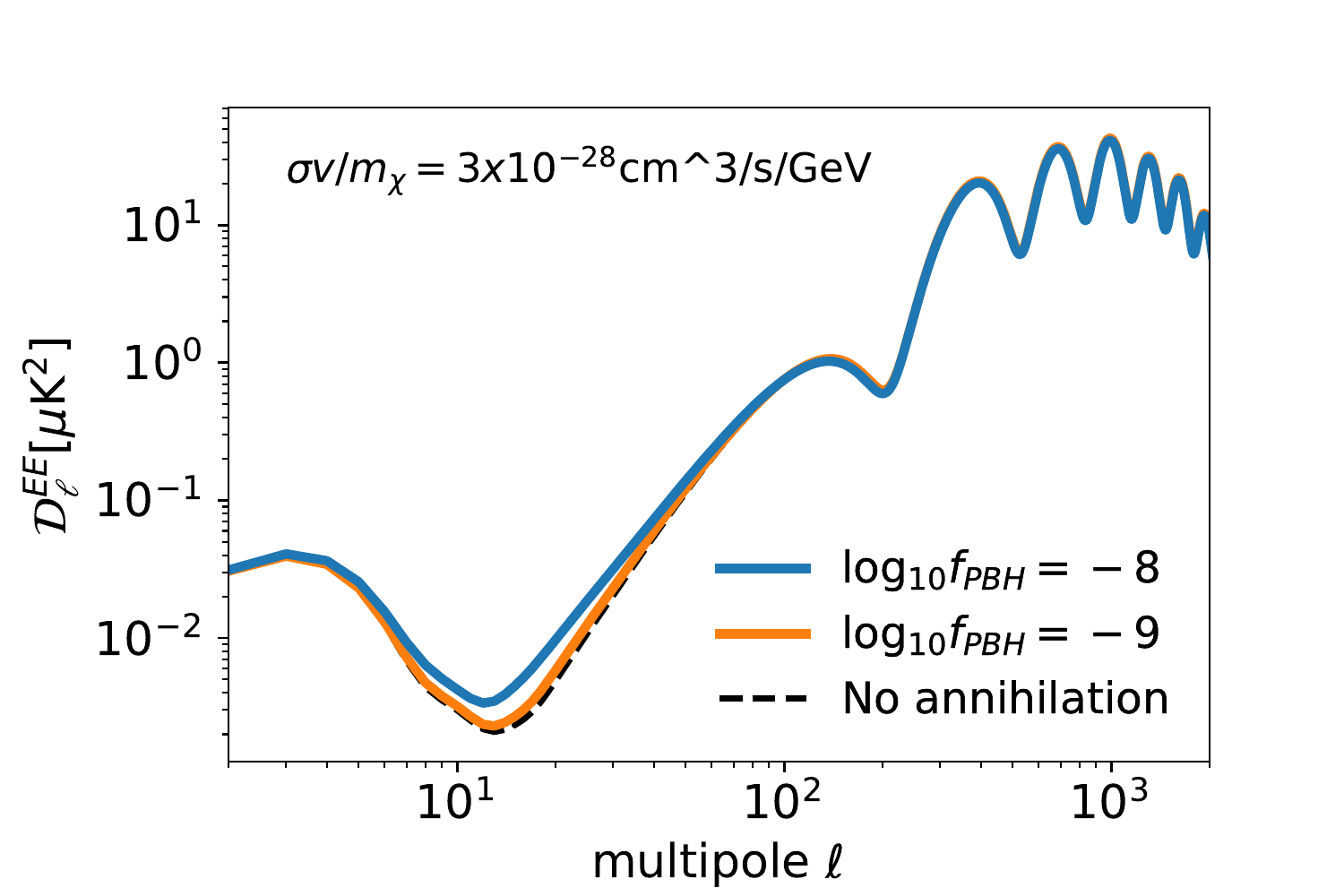}
  \caption{Angular power spectrum of CMB E-mode polarization. From the top to the bottom, the solid lines are the power spectra for
   $ f_{\rm PBH} = 10^{-8}$ and $ f_{\rm PBH} = 10^{-9}$. For reference, we show the angular power spectrum without the DM annihilation with a dashed line.}
  \label{fig:clee}
\end{center}
\end{figure}

We calculate the CMB anisotropies with the public Boltzmann code~{\tt CLASS}~\cite{2011JCAP...07..034B} modified to incorporate Eq.~\eqref{eq:dEdVdt}.
We plot the angular power spectra of the CMB E-mode polarization with different $f_{\rm PBH}$ in Fig.~\ref{fig:clee}.
In this figure, we assume the "tanh"-shaped reionization history and, according to the Planck best-fit cosmological parameters~\cite{2020A&A...641A...6P}, we set $z_{\rm reio} = 7.68$, where the ionization fraction becomes $x_e=0.5$ at $z_{\rm reio}$.
Large $f_{\rm PBH} $ induces the early reionization
and the enhancement of the optical depth from the last scattering surface.
As a result,
the modification arises
on the so-called reionization bump visible at $\ell \lesssim 20 $ in the CMB polarization angular power spectrum. Figure~\ref{fig:clee} shows that
the tail of the reionization bump is lifted as $f_{\rm PBH} $ becomes large.
We find that $f_{\rm PBH} = 10^{-9}$ can enhance the
amplitude of the reionization bump by $\sim 30$ percent.

\begin{figure}
  \begin{center}
  \includegraphics[width=10.0cm]{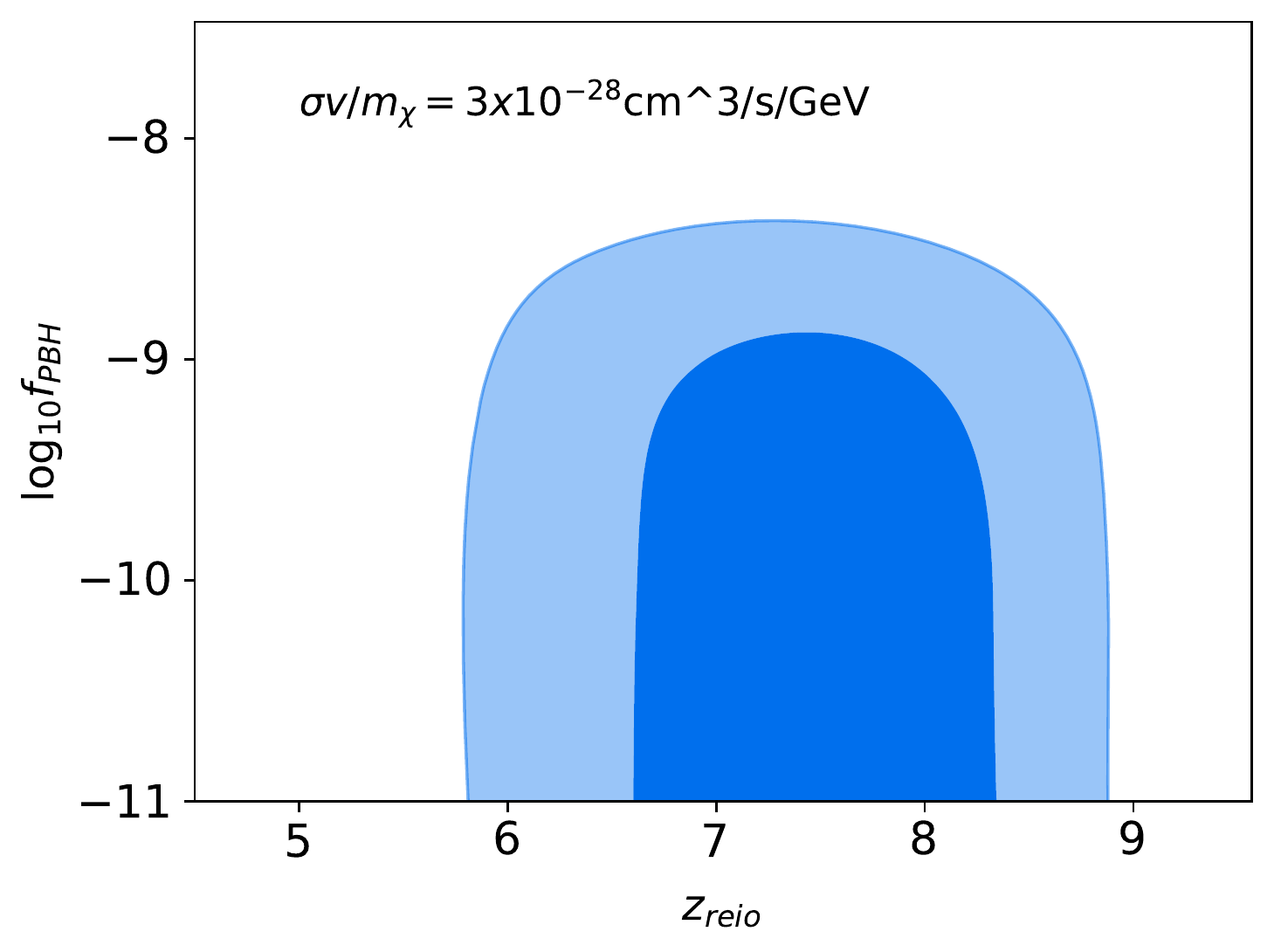}
  \caption{2D plot for $z_{\rm reio}$ and $f_{\rm PBH}$ from the MCMC analysis with Planck 2018 data. Here we set $\langle \sigma v \rangle /m_{\chi} = 3 \times 10^{-28} \rm  cm^3 /s/ GeV$.}
  \label{fig:zre_fpbh}
\end{center}
\end{figure}

In order to obtain the limit on $f_{\rm PBH }$ from CMB anisotropy observations,
we perform the MCMC analysis using {\tt Monte Python}~\cite{Audren:2012wb} with the modified {\tt CLASS}.
For CMB observation data, we use the baseline likelihood~(TTTEEE-lowl-lowE) from the 2018 data release~\cite{Aghanim:2019ame}.
The cosmological parameter set for the analysis is $\{\omega_b, \omega_d, 100\theta_s, \ln (10^{10}A_s), n_s, z_{\rm reio}, f_{\rm PBH}\}$.
We show the 2D contour plot for $z_{\rm reio}$ and $f_{\rm PBH}$ in Fig.~\ref{fig:zre_fpbh}.

We summarize our constraint on $f_{\rm PBH}$ for the~95\% confidence level in Table~\ref{tab:CMB_PBH}.
CMB constraints depend on the combination of the annihilation cross section~$\langle \sigma v \rangle $ and the mass~$m_\chi$,
and include the uncertain model parameter $f_{\rm ann}$.
When $  \langle \sigma v \rangle/m_{\chi} <3 \times 10^{-28} {\rm cm^3/s/GeV}$ with $f_{\rm PBH} ={\cal O}(10^{-9})$,
the boost factor given by Eq.~\eqref{eq:boost} is $B(z) \gg 1$ around $z \sim 20$. This means that we can neglect
the smooth-background DM contribution in~Eq.~\eqref{eq:dEdVdt} and it is enough to consider only the PBH halo contribution in these parameter regions.
Therefore, according to Eq.~\eqref{eq:injection_pbh}, the constraint on $f_{\rm PBH} $ can
be rewritten in
\be
f_{\rm ann}f_{\rm PBH} < 5.5 \times 10^{-9}
\left( \frac{ \langle \sigma v \rangle/m_{\chi}}{
3 \times 10^{-29} {\rm cm^3/s/GeV}}
\right)^{-1/3}.
\ee
This constraint is applicable when the PBH contribution dominates the smooth-background contribution [the explicit contribution ratio is given by Eq. \eqref{eq:boost}].

\begin{figure}
  \begin{center}
  \includegraphics[width=10.0cm]{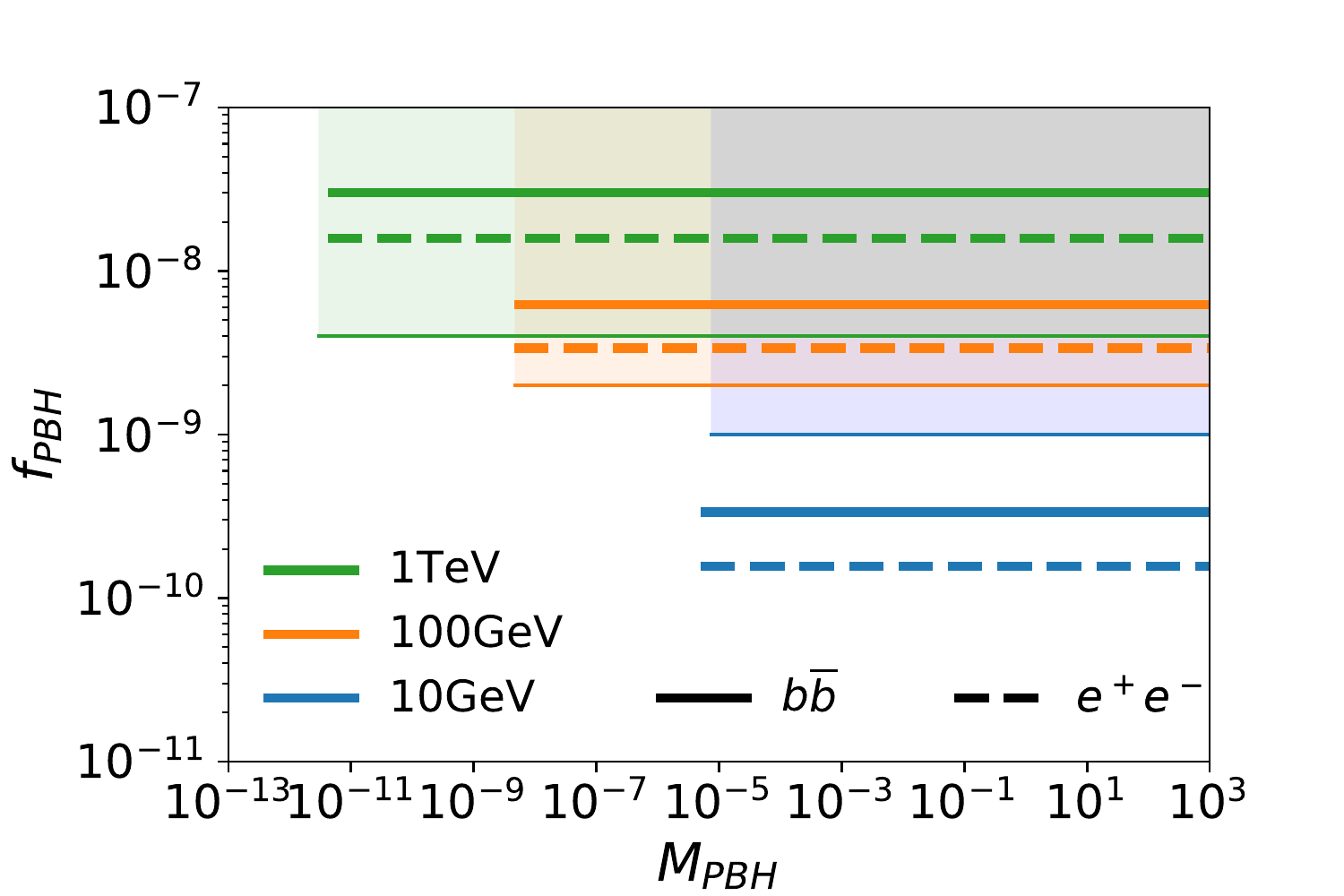}
  \caption{The constraint on $f_{\rm PBH}$ from Planck 2018 data for the canonical WIMP annihilation cross section $\langle \sigma v \rangle = 3 \times 10^{-26} \rm  cm^3 /s$. Here we consider that WIMPs annihilate into $b \bar b$ and $e^+e^-$ shown by the solid and dashed lines, respectively. From top to bottom, we set the WIMP mass $m_\chi =1~$TeV, 100~GeV, 10~GeV. The colored regions are ruled out by the extragalactic gamma-ray observations for the $b \bar b$ annihilation channel ~\cite{Adamek:2019gns}.}
  \label{fig:const_fpbh}
\end{center}
\end{figure}

The current Planck bound gives $\langle \sigma v\rangle/m_{\chi}\lesssim 3 \times 10^{-27} \rm cm^3 s^{-1}GeV^{-1}$ for the s-wave annihilating DM, and this in turn requires $m_\chi \gtrsim 10~{\rm GeV}$ for the canonical thermal WIMP cross section $\langle \sigma v \rangle = 3 \times 10^{-26}$~$\rm cm^3/s$ \cite{Slatyer:2015jla}.
Our constraint includes the uncertain model parameter $f_{\rm ann}$
which depends on the WIMP mass and the annihilation channel.
As shown in Fig.~\ref{fig:xe_temp} (and also as can be inferred by Eq. \eqref{eq:boost}),
the energy injection below $z \sim 500$ is important in our model.
To evaluate the impact of $f_{\rm ann}$ on the constraint,
we calculate the averaged $f_{\rm ann}$ in
$10< z < 500$ using the table of $f_{\rm ann}$ in Ref.~\cite{Slatyer:2009yq}
for the case of WIMPs annihilating into $e^+e^-$, $\mu^+\mu^-$ and $b\bar b$.
Using the averaged $f_{\rm ann}$,
we obtain the constraint for the canonical thermal WIMP scenario in Table~\ref{tab:CMB_PBH_canonical} and plot them in Fig.~\ref{fig:const_fpbh}
in solid lines for the $b\bar{b}$ and dashed lines for the $e^+e^-$ annihilation channels.
In previous works,
the PBH abundance with the annihilating WIMPs
is constrained by the observations of
Galactic and extragalactic background gamma-ray flux \cite{Lacki:2010zf,Boucenna:2017ghj,Adamek:2019gns,Eroshenko:2016yve,Carr:2020mqm,Cai:2020fnq,Delos:2018ueo,Kohri:2014lza,Bertone:2019vsk,Ando:2015qda,Hertzberg:2020kpm,Yang:2020zcu,Zhang:2010cj}.
Our constraints are much stronger than those from
the Galactic gamma-ray background and comparable with an
extragalactic one.
For comparison, we show the excluded regions
by extragalactic background in Ref.~\cite{Adamek:2019gns}
as the colored regions in Fig.~\ref{fig:const_fpbh}.
Comparing with Ref.~\cite{Adamek:2019gns}, the CMB anisotropy measurement provides a tighter constraint for small WIMP masses $m_\chi < 100$~GeV, while, for heavier masses $m_\chi > 100$~GeV, the constraint from the CMB anisotropy measurement is weaker than that from the extragalactic gamma-ray observations.

\begin{table}
  \begin{center}
    \begin{tabular}{|c|c|c|} \hline
	$\langle \sigma v \rangle /m_\chi
	~[\rm cm^3/s/GeV$]  	
	& ~$f_{\rm ann} f_{\rm PBH}$ (95 \%  C.L.) ~  \\\hline
      $3\times 10^{-29} $  & $ < 5.5 \times 10^{-9}$\\
$3\times 10^{-28}$ & $<  1.6 \times 10^{-9}$ \\
$3\times 10^{-27}$ & $ <  8.8 \times  10^{-11}$\\\hline
    \end{tabular}
  \caption{The constraints on $f_{\rm PBH}$ from Planck 2018 data for different values of $\langle \sigma v \rangle /m_\chi$.}
      \label{tab:CMB_PBH}
      \end{center}
\end{table}

\begin{table}
  \begin{center}
    \begin{tabular}{|c|c|c|c|c|} \hline
	$m_\chi$
	& ~$f_{\rm PBH}$ ($\chi \chi \rightarrow \gamma \gamma$) ~
	& ~$f_{\rm PBH}$ ($\chi \chi \rightarrow e^+e^-$) ~ &~$f_{\rm PBH}$ ($\chi \chi \rightarrow \mu^+ \mu^-$) ~&~$f_{\rm PBH}$ ($\chi \chi \rightarrow b \bar b$) ~   \\\hline
      $ 1~$TeV  & $ < 5.5 \times 10^{-9}$
      &$ <1.4 \times 10^{-8}$ & $ < 3.7 \times 10^{-8}$& $ < 2.7 \times 10^{-8}$ \\
$100~$GeV & $<  1.6 \times 10^{-9}$ &$< 3.6 \times 10^{-9}$& $< 9.4 \times 10^{-9}$
& $< 6.6 \times 10^{-9}$\\
~$10~$GeV~& $ <  8.8 \times  10^{-11}$ &$ < 1.4 \times 10^{-10}$& $ < 3.8 \times 10^{-10}$& $ < 3.0 \times 10^{-10}$\\\hline
    \end{tabular}
  \caption{The constraints on $f_{\rm PBH}$ in the canonical WIMP annihilation cross section, $\langle \sigma v \rangle=3 \times 10^{-26}~\rm cm^3/s$. Here we consider WIMPs annihilating into $\gamma \gamma$, $e^+e^-$, $\mu^+\mu^-$ and $b \bar b$.}
      \label{tab:CMB_PBH_canonical}
      \end{center}
\end{table}

\subsection{Global 21-cm signal}
\label{sec6}

The redshifted 21-cm signal depends on the thermal state of the intergalactic medium~(IGM). Therefore, the measurement of its signal from high redshifts can reveal the thermal history of the IGM.
Through the investigation of the effects of PBHs and WIMP annihilation on the thermal history,
the 21-cm signal is expected to provide a constraint on the abundance
of PBHs~\cite{2013MNRAS.435.3001T,2017JCAP...08..017G, 2019PhRvD.100d3540M} and the WIMP annihilation~\cite{2007MNRAS.377..245V,2009PhRvD..80d3529N} at high redshifts.
Recently Ref.~\cite{Yang:2020zcu} has studied the
impact of the WIMP annihilation on the global~(all-sky averaged) 21-cm signal in the mixed DM scenarios consisting of PBHs and a thermal canonical WIMP model with $m_\chi =100~$GeV.
Here we revisit the evolution of the global 21-cm signal in
the same mixed DM scenarios, covering a wider range of WIMP masses.

In 21-cm observations, the strength of the signal is measured in terms of so-called differential brightness temperature, which is the difference of the 21-cm brightness temperature
from the CMB one~(for a review, see Ref.~\cite{2006PhR...433..181F}).
The global differential brightness temperature from a redshift~$z$ is calculated  at~\cite{1997ApJ...475..429M,2010PhRvD..82b3006P}
\be
\label{eq:differential_temp}
\delta T_b (z)=
\frac{3}{32 \pi} \frac{hc^3 A_{10}}{k_B \nu_0^2} \frac{x_{\rm HI} n_H}{(1+z) H(z)} \left( 1-\frac{T_\gamma}{T_S}\right),
\ee
where
$A_{10}$ is the spontaneous emission coefficient of the 21-cm transition,~$A_{10} = 2.85 \times 10^{-15} \rm s^{-1}$, and
$T_s$ is the spin temperature of the neutral hydrogen hyperfine structure.
The spin temperature is determined by the balance
between the excitation and deexcitation in the hyperfine structure,
\be
T_S = \frac{T_\gamma + y_{\rm kin} T_k}{1+y_{\rm kin}},
\label{eq:spin}
\ee
where $y_{\rm kin}$ is the efficiency ratio between
the absorption of CMB photons and the thermal collisions
in the hyperfine transition.
In order to obtain $y_{\rm kin}$,
we take the approximated analytical form of $y_{\rm kin}$ in~Ref.~\citep{2006ApJ...637L...1K}.
In Eq.~\eqref{eq:differential_temp}, we ignore the contribution from Ly-$\alpha$ coupling,
because it becomes efficient after the formation of the first stars and galaxies.

\begin{figure}
 \begin{tabular}{cc}
 \begin{minipage}{0.45\hsize}
  \begin{center}
   \includegraphics[width=75mm]{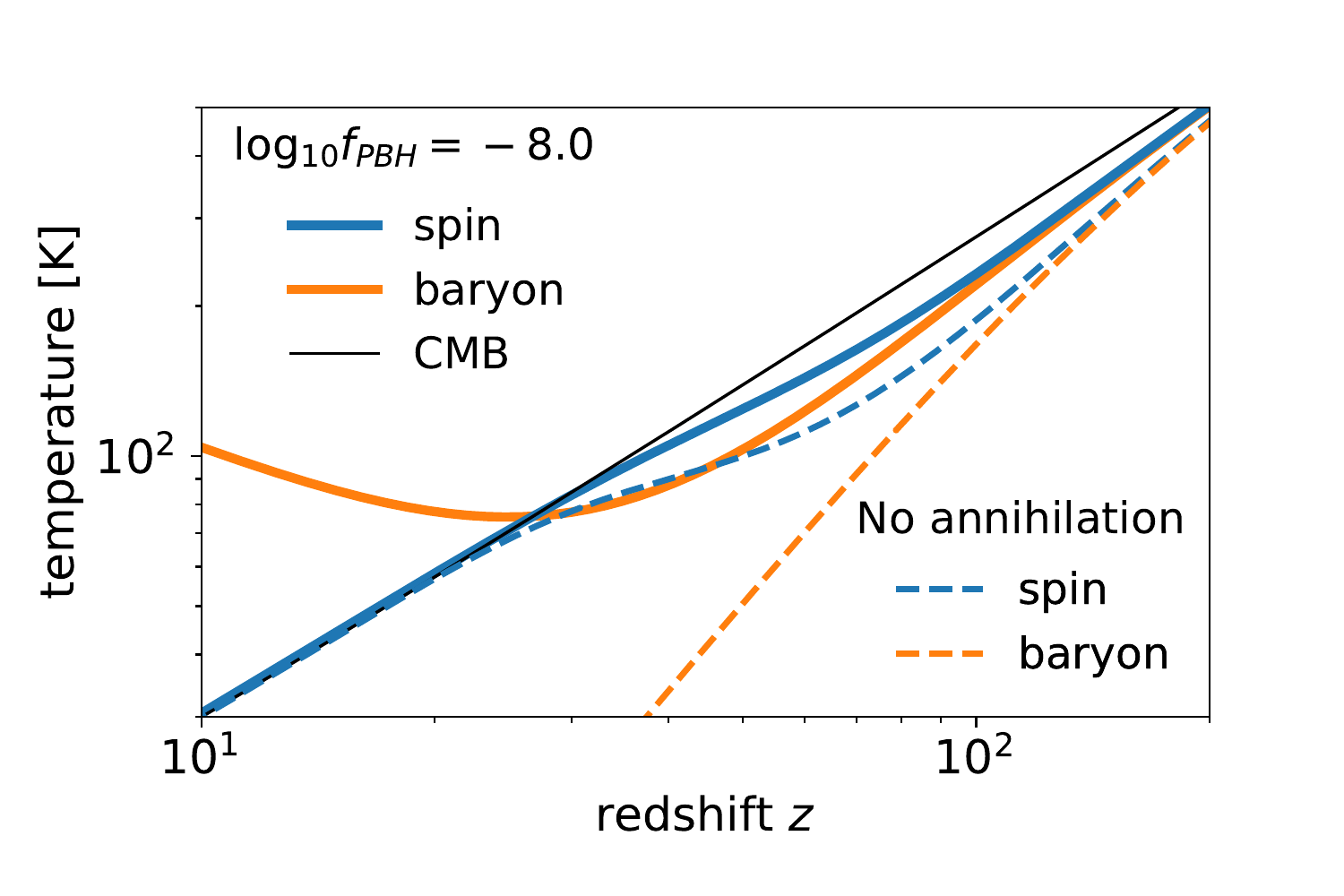}
  \end{center}
 \end{minipage}
 \begin{minipage}{0.45\hsize}
  \begin{center}
   \includegraphics[width=75mm]{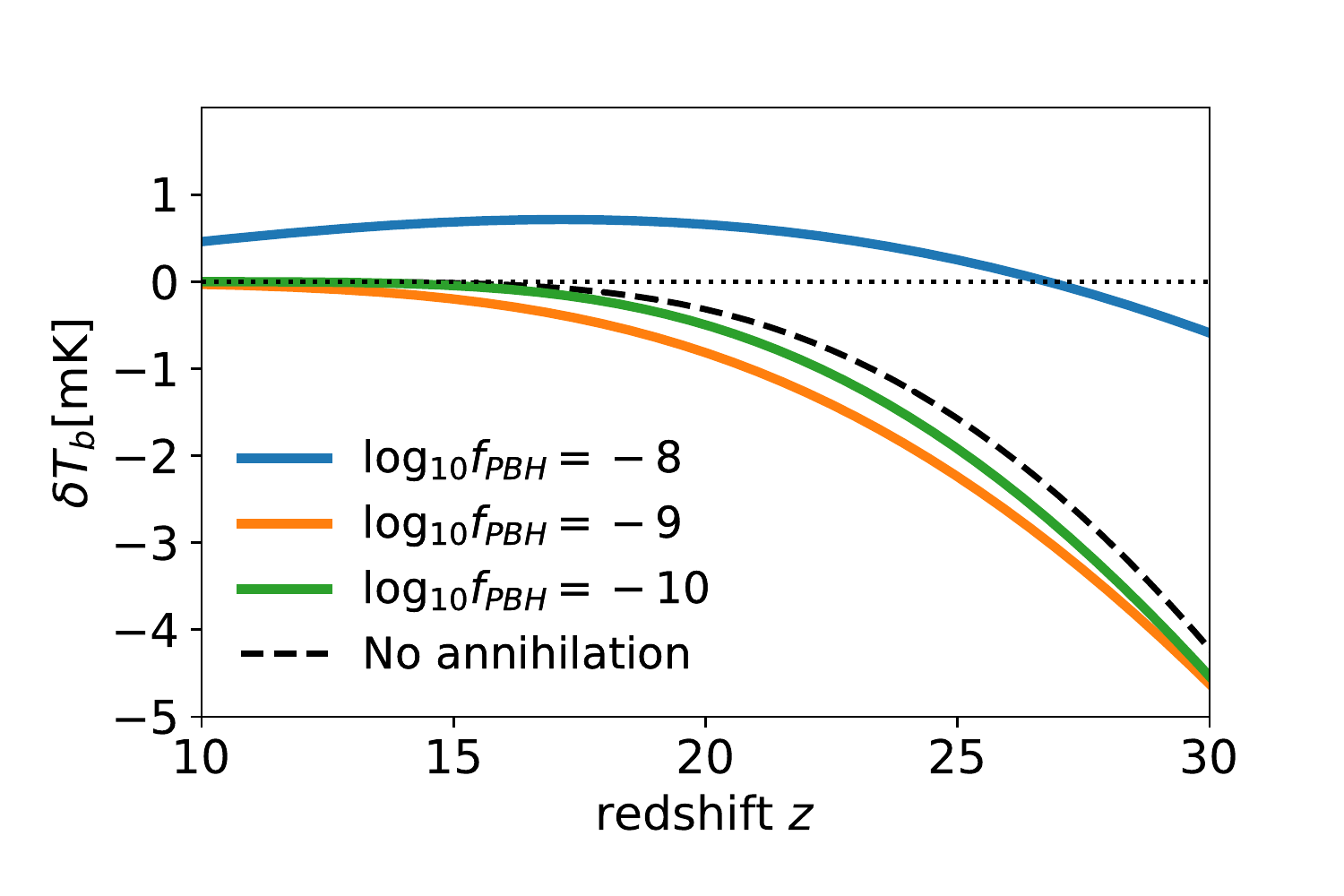}
  \end{center}
 \end{minipage}
 \end{tabular}
   \caption{{\it Left panel}: The spin and baryon temperature as functions of redshifts. The solid blue and orange lines represent the spin and baryon temperatures for $ f_{\rm PBH} = 10^{-8}$. For comparison, we plot their evolutions for no annihilation DM case in the dashed lines.
   {\it Right panel}:
   The evolution of the global differential brightness temperature with different $f_{\rm PBH}$.
   From top to bottom, the solid lines are for $ f_{\rm PBH} = 10^{-8}$, $f_{\rm PBH} =10^{ -9}$ and $ f_{\rm PBH} =10^{ -10}$. For reference, the dashed line shows the evolution in the case of no annihilation DM.
}
  \label{fig:dTb}
\end{figure}

The left panel of Fig.~\ref{fig:dTb} shows the effect of WIMP annihilation on the evolution of spin and baryon temperatures with
$ f_{\rm PBH} = 10^{-8}$ with $\langle \sigma v \rangle /m_\chi=3 \times 10^{-28}~[\rm cm^3/s/GeV]$ in the solid lines.
For comparison, we also give the evolutions in the standard cosmology~(no-annihilation) case using dashed lines.
WIMP annihilation heats up baryons and causes the baryon thermal evolution to deviate from the adiabatic evolution, $T_k \propto (1+z)^2$. Therefore, the spin temperature is also larger
than in the "no-annihilation" case.
The effective heating due to the annihilation can
make the baryon temperature exceed the CMB temperature.
In this case,
the spin temperature also becomes larger than the CMB temperature.

We plot the dependence of the
the global differential brightness temperature
on the PBH abundance in the right panel of Fig.~\ref{fig:dTb}.
A positive amplitude of the differential brightness temperature means an emission line on the CMB frequency spectrum
and a spin temperature is larger than the CMB temperature.
On the other hand, a negative one represents the
absorption line for CMB and a spin temperature smaller than the CMB temperature.
When $f_{\rm PBH}=10^{-8}$, the signal shifts from absorption to emission at $z_{\rm tr} \approx 27$.
As $f_{\rm PBH}$ decreases, the annihilation effects become inefficient. As a result, the transition redshift, $z_{\rm tr}$, also becomes small.
We found that, if $f_{\rm PBH}<2 \times 10^{-9}$, the baryon temperature cannot exceed the CMB temperature before the epoch of reionization,~($z>7$). In the case of $f_{\rm PBH}<2 \times 10^{-9}$, the sign of the signal is always negative.
When $f_{\rm PBH}$ becomes smaller, the evolution of the signal approaches the one in the "no-annihilation" case.

\begin{figure}
  \begin{center}
  \includegraphics[width=10.0cm]{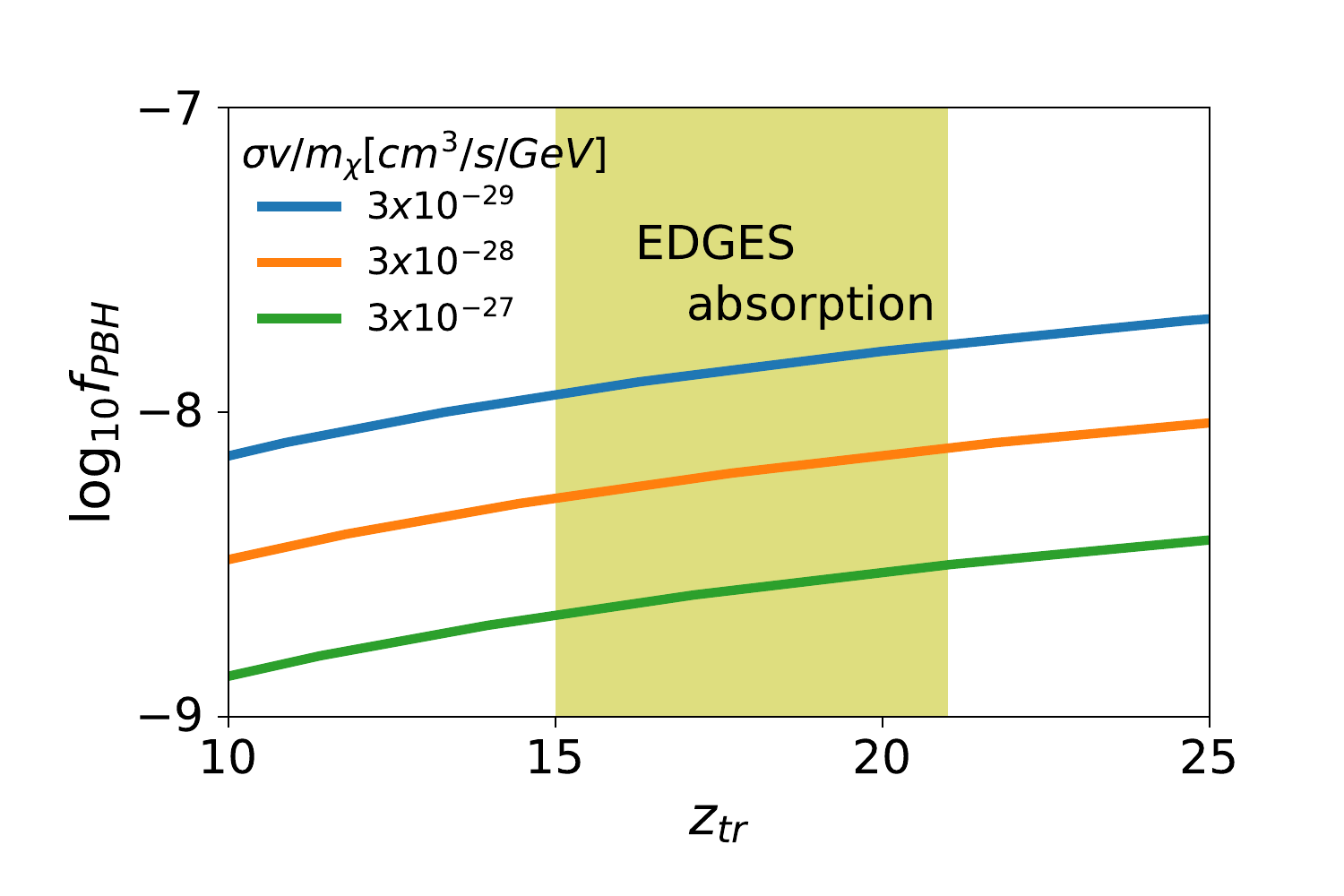}
  \caption{The dependence of the transition redshift,~$z_{\rm tr}$, from the absorption to emission on the PBH fraction $f_{\rm PBH}$.}
  \label{fig:edges-21cm}
\end{center}
\end{figure}

Recently, EDGES reported that they have detected the global absorption signals of redshifted  21-cm lines
from the redshift range between $z\sim 21$ and $z\sim 15$~\cite{2018Natur.555...67B}.
If this measurement is confirmed, the baryon temperature is lower than the CMB temperature until $z\sim 15$. Therefore, this measurement can provide a constraint on the heating source.
In Fig.~\ref{fig:edges-21cm}, we represent
the relation between $z_{\rm tr}$ and $f_{\rm PBH}$. Small $f_{\rm PBH}$ provides low $z_{\rm tr}$.
Therefore, from this figure,
we can conclude that
the EDGES absorption signal
at $z\sim 15$ can give the limit on the PBH abundance of
$f_{\rm PBH} < {\cal O}(10^{-9})$ for $10~{\rm GeV} < m_{\chi} < 1~{\rm TeV}$ with the canonical thermal WIMP cross section $\langle \sigma v \rangle= 3\times 10^{-26} \rm cm^3/s$.

Our constraint is weaker than the one in Ref.~\cite{Yang:2020zcu}.
To obtain our constraint,
we take the criteria for the constraint which
is the transition redshift from the absorption to the emission.
The amplitude of the signals detected by EDGES is $500^{+500}_{-200}~\rm mK$~\cite{2018Natur.555...67B}. Taking this at face value, Ref.~\cite{Yang:2020zcu} requires
the condition that the differential brightness temperature
is less than $\delta T_{\rm b} <-100~\rm mK$.
In order to obtain such a large absorption signal,
the contribution of the Ly-$\alpha$
coupling, which we ignore in Eq.~\eqref{eq:spin}, is required.
The efficiency of the coupling depends on the formation history of luminous objects in the Universe and has a large theoretical uncertainty.
Assuming the strong Ly-$\alpha$ coupling, $T_{S} = T_{k}$,
we calculate the 21-cm signal evolution again.
When we adopt the criteria, $\delta T_{\rm b} <100~\rm mK$, during the EDGES observation redshifts,
we obtain the constraints, $f_{\rm PBH} < 4 \times 10^{-10}$ and $<2 \times 10^{-9}$ for $m_\chi = 100~$GeV and $1~$TeV with $\langle \sigma v \rangle = 3 \times 10^{-26} \rm  cm^3 /s$.
Therefore, 21-cm global observations have a potential to provide as tight constraints as extragalactic gamma-ray observation even for a large-mass WIMP.
Note that the criterion, $\delta T_{\rm b} <-100~\rm mK$, can exclude the canonical thermal WIMP cross section for $m_\chi < 100~$GeV~\cite{2019PhLB..789..137C}.

\section{conclusion}
\label{secconc}

In this paper, we have investigated the constraint on the PBH abundance in the presence of self-annihilating WIMPs.
In previous works, the stringent constraint on the PBH in this mixed dark matter scenario has been obtained from the Galactic and extragalactic gamma-ray data.
We focused on the CMB anisotropy measurements and global 21-cm observations.
If PBHs exist, the steep WIMP DM halo is created around a PBH and the annihilation is enhanced inside such a DM halo.
The energy released in the annihilation can heat and reionize diffuse background baryon gas outside DM halos
and modify the thermal history of diffuse baryon gas.
This modification can make a deviation in the CMB anisotropies and 21-cm global signals from those in the standard $\Lambda$CDM model.

In order to obtain the constraint from the CMB measurement, we have studied the effect of
the WIMP annihilation from dark matter halos around PBHs
and performed the MCMC analysis with the latest Planck data.
Our constraints from the Planck data, for $b\bar{b}$ annihilation channel examples, are
$f_{\rm PBH} \lesssim 3\times 10^{-10},  7\times 10^{-9}$, and $3\times 10^{-8}$ respectively for $m_{\chi}=$ 10~GeV, 100~GeV and 1~TeV, respectively, for the canonical thermal annihilation cross section $\langle \sigma v \rangle=3\times 10^{-26} \rm cm^3/s$.
These bounds are stronger than the limits from the Galactic gamma-ray background
and comparable with the one from the extragalactic gamma-ray background.
Not all of energy produced in the annihilation can be absorbed into
baryon gas. The efficiency depends on the WIMP mass and the annihilation channels. We have found that, in the mixed DM scenario with PBHs and WIMPs, the thermal history of baryon gas is sensitive to the efficiency in the redshift range $10<z<500$.
In this paper, we have presented the constraints for the annihilation into $\gamma \gamma$, $e^+e^-$, $\mu^+\mu^-$ and $b \bar b$.
Using the averaged efficiency factor in this redshift range, one can easily convert our constraint to the one in the different annihilation channels.

Our studies focused on the parameter space for which the UCMH around PBH was numerically verified to possess the steep density profile $\rho(r) \propto r^{-9/4}$. The dark matter profile around the PBH where the DM velocity dispersion cannot be ignored is heavily model dependent (e.g. on the DM properties such as the nature of DM kinetic decoupling \cite{Loeb:2005pm,Bertschinger:2006nq,Gondolo:2012vh, Profumo:2006bv,Gondolo:2016mrz,Green:2003un,Green:2005fa,Bringmann:2006mu}) and the dedicated numerical simulations have not been performed yet even though the analytical estimation has been done
assuming the Maxwell–Boltzmann distribution for the DM velocity \cite{Eroshenko:2016yve,Boucenna:2017ghj}. The bounds on $f_{PBH}$ become much weaker for those less steep DM profiles in existence of the nonradial motion of DM bound to the PBH \cite{Carr:2020mqm}. More detailed numerical studies where one needs to account for the DM velocity distributions in their accretion onto the PBHs are left for future work.

We have also studied the constraint from the global 21-cm signals.
Before the EoR, the global 21-cm signals are predicted as  absorption signals on the CMB frequency spectrum. However the baryon gas heating by the WIMP annihilation can shift 21-cm signals from absorption to emission.
Recently the EDGES experiment has reported the detection of the absorption signals from the redshifts, $15 \lesssim  z \lesssim 25$.
Motivated by the EDGES report, we have adopted the criterion that the global 21-cm signals cannot turn into the emission until $z<15$.
The obtained constraint is $f_{\rm PBH} < {\cal O}(10^{-9})$ for $10~{\rm GeV} < m_{\chi} < 1~{\rm TeV}$ with $\langle \sigma v \rangle=3 \times 10^{-26}~\rm cm^3/s $.
Although this constraint is slightly weaker than the CMB constraint,
we have demonstrated that further development in both theory and observation of the 21-cm signals before the EoR can provide a stringent constraint which is better than the limit from the extragalactic gamma-ray background.
Assuming the strong Ly-$\alpha$ coupling limit and the criterion,
$\delta T_{\rm b} <100~\rm mK$,
the constraints are improved to $f_{\rm PBH} < 4 \times 10^{-10}$ and $<2 \times 10^{-9}$ for $m_\chi = 100~$GeV and $1~$TeV with the canonical WIMP annihilation cross section $\langle \sigma v \rangle=3 \times 10^{-26}~\rm cm^3/s $.

\begin{acknowledgments}
This work was supported by the Institute for Basic Science (IBS-R018-D1) and Grants-in-Aid for Scientific Research from JSPS (21K03533).
\end{acknowledgments}



\end{document}